\documentclass{PoS}

\usepackage{amssymb,latexsym}
\usepackage{amsmath}
\usepackage{multirow}
\usepackage{amsfonts,slashed}
\usepackage{bm,euscript,array,cite}
\usepackage[dvips]{epsfig}

\title{Fuzzy extra dimensions and particle physics models}

\ShortTitle{Fuzzy extra dimensions and particle physics models}

\author{\speaker{Athanasios Chatzistavrakidis}
\\
        Bethe Center for Theoretical Physics and Physikalisches Institut, Universit\"{a}t Bonn\\ 
	Nussallee 12, D-53115 Bonn, Germany \\
        E-mail: \email{than@th.physik.uni-bonn.de}}

\author{Harold Steinacker\\
   Department of Physics, University of Vienna\\
Boltzmanngasse 5, A-1090 Wien, Austria     \\
        E-mail: \email{harold.steinacker@univie.ac.at}}

\author{George Zoupanos\footnote{On leave from Physics Department, National
Technical University of Athens, Zografou Campus, GR-15780 Zografou, Greece.}\\
   Institute for Theoretical Physics, University of Heidelberg \\
Philosophenweg 16, D-69120 Heidelberg, Germany\\
        E-mail: \email{George.Zoupanos@cern.ch}}

\abstract{In the present contribution the construction of particle physics models in theories with fuzzy extra dimensions is discussed. 
We focus on a bottom-up approach where the structure of a higher-dimensional theory arises within ordinary 
four-dimensional field theory via an appropriate mechanism based on spontaneous symmetry breaking. 
Subsequently, possible ways to obtain particle physics models with realistic spectrum 
in this framework are exhibited.}

\FullConference{Corfu Summer Institute on Elementary Particles and Physics - Workshop on Non 			Commutative Field Theory and Gravity,\\
		September 8-12, 2010\\
		Corfu Greece}

\newcommand{\be}{\begin{equation}} \newcommand{\ee}{\end{equation}}
\newcommand{\beq}{\begin{equation}} \newcommand{\eeq}{\end{equation}}
\newcommand{\beqa}{\begin{eqnarray}}
\newcommand{\eeqa}{\end{eqnarray}} \newcommand{\eq}[1]{(\ref{#1})}
\def\nn{\nonumber} \def\bea{\begin{eqnarray}} \def\eea{\end{eqnarray}}
\def\obar{\overline}

\newcommand{\barr}{\begin{array}}
\newcommand{\earr}{\end{array}}

\def\a{\alpha}  
 \def\g{\gamma} \def\G{\Gamma}
  
 \def\e{\epsilon} 
\def\f{\phi} \def\F{\Phi}   
\def\l{\lambda} \def\L{\Lambda} \def\la{\lambda} 
 \def\o{\omega}

      \def\cN{{\cal N}}

\def\R{{\mathbb R}} \def\C{{\mathbb C}} 
\def\Z{{\mathbb Z}} \def\one{\mbox{1 \kern-.59em {\rm l}}}

\def\beq{\begin{equation}}
\def\eeq{\end{equation}}
\def\bea{\begin{eqnarray}}
\def\eea{\end{eqnarray}}

\def\g{\gamma}
\def\1{\frac{G_2}{SU(3)}}
\def\2{\frac{Sp_4}{SU(2)\times U(1)}}
\def\3{\frac{SU(3)}{U(1)\times U(1)}}
\def\bi{\begin{itemize}}
\def\ei{\end{itemize}}

\def\l{\lambda}
\def\a{\alpha}

\def\diag{{\rm diag}}

\begin{document}

\section{Introduction}

The aim of unifying all the observed interactions has always been in the center of interest within the theoretical physics community. The interest is not based only on obvious easthetic reasons but is deeply related to the fact that the Standard Model (SM) of
elementary particle physics contains over twenty free parameters. In addition, its minimal supersymmetric extension (MSSM), which is the the best bet of many particle physicists as the model that will be able to describe the physics beyond the SM, has around a hundred more
parameters. It is then natural to expect that a truly unified picture of nature will be able to reduce tremendously the number of free parameters of the particle physics models.
 One of the most exciting approaches to the unification quest is based on the assumption that there exist extra dimensions beyond the four that have been observed. This proposal has obviously its roots in the old Kaluza-Klein observation that the reduction of
five-dimensional gravity can lead to electromagnetism coupled to gravity. In recent years a strong theoretical support of the existence of extra dimensions comes from superstring theories, which are very serious candidates for a unified description of all fundamental
interactions including gravity; superstring theories can be consistently defined only in higher than four dimensions.
Finally the possibility that the inverse size of the extra dimensions can consistently be of the order of TeV \cite{Antoniadis:1990ew} and therefore their existence becomes testable in current and future colliders certainly gave a huge boost in the popularity of this idea in a much
wider physics audience.

 Another framework aiming to describe both physics at the Planck scale as well as particle physics models is non-commutative geometry.  In recent years strings and noncommutatvity turned out to be closely related. For instance Seiberg and Witten \cite{Seiberg:1999vs} made the observation that
a natural realization of non-commutativity of space emerges by considering D-branes,
defined in type II superstring theory, in the presence of a constant antisymmetric field. Moroever the type IIB superstring theory, which is expected to be related to the other superstring theories by duality transformations \cite{Hull:1994ys}, in its non-perturbative formulation as a
matrix model \cite{Ishibashi:1996xs} is a non-commutative theory.
 
In the spirit of the above discussion, we investigate higher-dimensional gauge theories with non-commutative extra dimensions and their dimensional reduction in four dimensions. An up to-date overview of certain attempts in this direction, developed over the
last years can be found in \cite{Chatzistavrakidis:2010tq}. The development of these ideas has followed two complementary directions, namely (i) the dimensional reduction of a higher-dimensional gauge theory over non-commutative and in particular fuzzy internal spaces \cite{Aschieri:2003vy} 
and  (ii) the dynamical generation of fuzzy extra dimensions within four-dimensional and renormalizable gauge theories \cite{Aschieri:2006uw}. Due to lack of space we limit ourselves here in the second approach.
Therefore in the present article instead of reducing to four dimensions a higher-dimensional theory with fuzzy extra dimensions we reverse the problem and examine how a four-dimensional gauge theory develops fuzzy dimensions due to its spontaneous symmetry breaking. In addition we address the important problem of chirality in this framework, which has been recently succesfully resolved \cite{Chatzistavrakidis:2010xi}. Finally we present suggestive hints on how to construct fully phenomenologically viable particle models. 


\section{Dynamical generation of fuzzy extra dimensions}

A very illustrative example of the mechanism leading to the emergence of fuzzy extra 
dimensions, which was mentioned in the introduction, is provided by considering as a starting point a ${\cal N}=4$ SYM theory in 4D \cite{Brink:1976bc}. 
Therefore let us start by briefly discussing those features of the theory which are necessary for our purposes. 

A ${\cal N}=4$ supersymmetric $SU(N)$ gauge theory contains, in ${\cal N}=1$ language, a $SU(N)$ vector supermultiplet and three adjoint chiral supermultiplets $\F^i,i=1,2,3$. The component fields are the $SU(N)$ gauge bosons $A_{\mu},\mu=1,\dots ,4$, six adjoint real scalars
$\f^a,a=1,\dots, 6$, transforming as $\textbf{6}$ under the $SU(4)_R$ $R$-symmetry of the theory and four adjoint Weyl fermions $\psi^p,p=1,\dots ,4,$ transforming as $\textbf{4}$ under the $SU(4)_R$. The theory is defined on the Minkowski spacetime, whose coordinates are denoted as $x^{\mu}$.
The corresponding action is given by (spinor indices are supressed)
\bea
{\cal S}_{YM} &=& \int d^4 x\Biggl[Tr\biggl(-\frac 1{4} F_{\mu\nu}F^{\mu\nu}
+ \frac 12 \sum\limits_{a=1}^6\,D^\mu\phi^a D_\mu \phi^a
- V(\phi)\biggl)  \nn\\
&& \quad + \frac 12 Tr \Big(i\bar\psi \slashed D \psi
  + g \Delta_R^a\,\bar\psi R [\phi^a,\psi]
  - g \Delta_L^a\,\bar\psi L [\phi^a,\psi] \Big)\Biggl],
\label{YM-action-4D}
\eea where the potential has the form \beq \label{4dpot}
V(\phi) = V_{{\cal N}=4}(\phi)=-\frac{1}{4}g^2 \sum\limits_{a,b}
\, [\phi^a,\phi^b]^2 . \eeq 
In the above expressions $D_\mu = \partial_\mu - i g [A_\mu,.]$ is the four-dimensional 
covariant derivative in the adjoint representation. The projection operators
 $L$ and $R$ are, as usual, defined as $L= \frac 12 (\one - \gamma_5)$ and $R= 
\frac 12 (\one + \gamma_5)$. The
$\Delta_{L}^a$ and $\Delta_{R}^a$ are the intertwiners of
the $\mathbf{4}\times\mathbf{4} 
\to \mathbf{6}$ and $\mathbf{\bar 4}\times\mathbf{\bar 4} \to \mathbf{6}$ 
respectively, namely they are Clebsch-Gordan coefficients that couple two 
$\mathbf{4}$s to a $\mathbf{6}$. The Yukawa interactions in (\ref{YM-action-4D}) 
are separately invariant
under the $SU(4)$, since the $R \psi^p$ transforms in the $\mathbf{4}$
and the $L\psi^p$ in the $\mathbf{\bar 4}$ of the $SU(4)$.

The scalar potential (\ref{4dpot}) is positive definite and its minimum is at $[\f^a,\f^b]=0$. Obviously there is no room for 
non-commutative vacuum configurations without any further modifications. Therefore let us consider deformations of the theory, 
by adding to its scalar potential appropriate mass and trilinear terms.
Such terms do not respect the $R$-symmetry, 
i.e. they break the ${\cal N}=4$ supersymmetry. 
We can either restrict ourselves to soft supersymmetry breaking terms, or 
allow also more general (marginal) terms in the potential.
Let us focus on potentials which break the global $SU(4)\approx SO(6)$ symmetry down to $SO(3) \times SO(3)$.
Then the most general such  potential can be written (after suitable redefinitions) in the following form
\beq
V[\phi] = a_1^2 (\phi_i\phi_i + b_1\one)^2
 + a_2^2 (\phi_{i+3}\phi_{i+3} + b_2\one)^2
 + \frac 1{g_1^2} F_{ij} F_{ij} + \frac 1{g_2^2} F_{i+3,j+3} F_{i+3,j+3}
 + \frac 1{g_3^2}  [\phi_i,\phi_{j+3}][\phi_i,\phi_{j+3}] \label{V-S2S2}
\eeq 
where $i,j = 1,2,3$ and
\bea
F_{ij} &=& [\phi_i,\phi_j] - i\a_1\varepsilon_{ijk} \phi_k, \nn\\
F_{i+3,j+3} &=& [\phi_{i+3},\phi_{j+3}] - i\a_2\varepsilon_{ijk} \phi_{k+3} .
\eea
For suitable parameters 
$a_{1/2}, b_{1/2}, g_{1/2/3}$, it admits a stable global minimum given by the following relations, 
\bea
\, [\phi_i,\phi_j] &=& i \a_1\, \varepsilon_{ijk} \phi_k,\nn\\ \phi_i 
\phi_i &=& \a_1^2 \frac{N_1^2-1}4,
\label{fuzzysphere}
\eea
and similarly for the $\phi_{i+3}$, along with the mutual commutation relation \be  [\phi_i,\phi_{j+3}] = 0. \ee
It is easy to find a configuration satisfying the above relations,
\bea
 \phi_i \,\,\,\,\,\,&=&\,\, \a_1\, \la_i^{(N_1)}\otimes \one_{N_2}
\otimes\one_{n},\nn\\
 \phi_{i+3} \,\, &=& \,\,\a_2\, \one_{N_1}\otimes\la_{i}^{(N_2)}
\otimes\one_{n}, \qquad i=1,2,3,
\label{vacuum-typeI-S2S2}
\eea
where $\la_i^{(N)}$ denotes the generator of the
$N$-dimensional irreducible representation of $SU(2)$ and
\be
N = N_1 N_2\, n .
\label{typeI-condition}
\ee
Such a vacuum should be interpreted as a stack of $n$
fuzzy branes with geometry $S_{N_1}^2 \times S^2_{N_2}$ and in the present construction it breaks the
gauge group $SU(N)$ down to $SU(n)$.

Apart from the above configuration there is a more generic case where the vacuum has the form
\bea
\phi_i
&=& \a_1\left(\begin{array}{cc}\, \la_i^{(N_{11})}\otimes \one_{N_{21}}
\otimes\one_{n_1} & 0 \\ 0
& \,\la_i^{(N_{12})}\otimes \one_{N_{22}}\otimes\one_{n_2}
             \end{array}\right),  \nn\\
\phi_{3+i}
&=& \a_2\left(\begin{array}{cc}\, \one_{N_{11}}\otimes\la_{i}^{(N_{21})}
\otimes\one_{n_1} & 0 \\ 0
& \,\one_{N_{12}}\otimes\la_{i}^{(N_{22})}\otimes\one_{n_2}
             \end{array}\right),\quad i=1,2,3.  \nn\\
\label{vacuum-mod2_s2s2}
\eea
The commutant of the generators involved in the above configuration, i.e. the
unbroken gauge group, is
$SU(n_1) \times SU(n_2) \times U(1)_Q$, where the $U(1)_Q$
has generator
\be
Q = \left(\begin{array}{cc} \frac 1{N_{11} N_{21} n_1}\one & 0 \\ 0
& -\frac 1{N_{12} N_{22} n_2}\one
             \end{array}\right).
\ee
This vacuum corresponds to a splitting
\be
 N = n_1 N_{11} N_{21} + n_2 N_{12} N_{22},
\ee
which is more generic than \eq{typeI-condition}. The interpretation of this vacuum is as a stack of
$n_1$ fuzzy branes
and a stack of $n_2$ fuzzy branes
 with geometry $S^2_1 \times S^2_2$.
However, these fuzzy spheres carry
magnetic flux under the unbroken $U(1)_Q$
given by \cite{Steinacker:2003sd}
\be
m_1= N_{11}-N_{12}, \qquad m_2 = N_{21}-N_{22},
\ee
on $S^2_1$ and $S^2_2$ respectively.

According to the above, in the spontaneously broken phase the theory supports gauge groups of the types $SU(n)$ and $SU(n_1)\times SU(n_2)\times U(1)$. These gauge groups are obviously very appealing for phenomenological applications. However, the gauge group structure is of course not enough. In the following we shall also investigate the properties of the matter fields, concentrating on the possibility to obtain chiral fermions. 

Before proceeding, let us discuss in more detail some features of the resulting theory. As we saw above, the vacuum of the theory is obtained when the scalar fields acquire non-vanishing vevs given by eqs. (\ref{vacuum-typeI-S2S2}), (\ref{vacuum-mod2_s2s2}). These vevs satisfy the relations (\ref{fuzzysphere}), which define the fuzzy sphere.  The
fluctuations around this vacuum, $\phi_a + A_a$, provide the
components $A_a$ of a higher-dimensional gauge field $A_M = (A_\mu,
A_a)$, and the action can be interpreted as Yang-Mills theory on the
8-dimensional space $M^4 \times S^2\times S^2$, with gauge group depending
on the particular vacuum.  We therefore interpret the vacuum as
describing dynamically generated extra dimensions in the form of a product of two
fuzzy spheres $S^2_{N}\times S^2_N$. This geometrical interpretation can be fully
justified
 by working out the spectrum of Kaluza-Klein
modes \cite{Aschieri:2006uw}. The effective low-energy theory is then given by the zero
modes on $S^2_{N}\times S^2_N$. Let us also mention that the above procedure may be repeated in order to generate different fuzzy spaces by choosing appropriate deformations and parameters and studying the vacua of the resulting potentials.

Up to now we dealt only with the scalar sector of the theory. Let us now proceed to the study of the fermionic sector aiming at the description of massless fermions in the above context. Here we adopt a rather descriptive approach; more technical details may be found in \cite{Steinacker:2007ay,Chatzistavrakidis:2009ix}. Naturally, a vacuum of the type (\ref{vacuum-mod2_s2s2}) is expected to accommodate massless fermions protected by chirality due to the index theorem, since it includes magnetic fluxes. Therefore let us concentrate on this vacuum, where the fermionic wavefunction can be split as \be
\psi = \left(\begin{array}{cc} \psi^{11} & \psi^{12} \\
                              \psi^{21} & \psi^{22}
\end{array}\right).
\ee
The components $\psi^{11}$ and $\psi^{22}$ transform in the adjoint representations $(n_1^2-1,1)$ and $(1,n_2^2-1)$ of the non-abelian part $SU(n_1) \times SU(n_2)$ of the gauge symmetry respectively. Moreover, the component $\psi^{12}$ transforms in the bifundamental representation $(n_1,\obar n_2)$ of the
$SU(n_1) \times SU(n_2)$ and the component
$\psi^{21}$  in the $(\obar n_1,n_2)$. Accordingly, the diagonal components $\psi^{11}$ and
$\psi^{22}$ are unaffected by the magnetic flux while the
off-diagonal components $\psi^{12}$ and
$\psi^{21}$ feel this magnetic flux and develop chiral zero
modes according to the index theorem. Even though the fluxes on $S^2\times S^2$ lead indeed
to the expected zero modes, the model nevertheless
turns out to be non-chiral.
More precisely, we find essentially mirror models, where 
two chiral sectors arise with opposite chirality. 
This means that each would-be zero mode from $\psi^{12}$ 
has a mirror partner from $\psi^{21}$, with opposite
chirality and gauge quantum numbers. While this may still be physically interesting since the mirror fermions 
may have larger mass than the ones we see at low energies, it is more desirable to find a chiral version. This issue is addressed in the following section.
 
\newpage

\section{Fuzzy orbifolds and chirality}

The mechanism we described in the previous section provides a simple yet powerful way of obtaining particle physics models with 
promising gauge group structure. However, without any further requirements it cannot succeed in fulfilling one of the 
central requirements of a successful model, which is the chiral character of the weak force and the existence of chiral fermions. 
There are several different possibilities which lead to 
mechanisms which solve the above problem. A collection of such mechanisms in string/M-theory-inspired model building 
may be found in \cite{Uranga:2002vk}.

One of the simplest known mechanisms in order to generate chiral fermions is to consider D-branes at orbifold singularities 
\cite{Douglas:1997de}. Due to the close relation between D-branes and fuzzy spaces \cite{Myers:1999ps},
this mechanism turns out to be applicable also in the context of fuzzy extra dimensions.
Moreover, orbifolds provide a simple way to break the large amount of supersymmetry 
down to ${\cal N}=1$ \cite{Dixon:1985jw}. Before proceeding, let us stress 
that we are considering 
field-theoretic orbifolds and not string-theoretic ones. They are defined in terms of perfectly well-defined gauge theories with 
a particular structure, without any twisted sectors. 
Following \cite{Kachru:1998ys}, this is achieved by projecting  
the 4D theory (here ${\cal N}=4$ SYM) under the action of a discrete symmetry group $\G$. This
corresponds to
an orbifold compactification of a 10D theory (usually a type II string theory) on $\C^3/\G$.

With this in mind,
let us consider again a $SU(M)$ ${\cal N}=4$ SYM theory. For later convenience the integer $M$ is taken to be a multiple of 3, 
namely $M=3N$, and therefore the gauge group is $SU(3N)$. In order to perform an orbifold projection of this theory we have to 
consider the discrete group $\G$ as a subgroup of $SU(4)_R$, the $R$-symmetry of the theory.
 There are three possibilities, which have a direct impact on the amount of remnant supersymmetry \cite{Kachru:1998ys}:
\begin{enumerate} \item $\G$ is maximally embedded in $SU(4)_R$, in which case we are generically led to models without supersymmetry. 
\item $\G$ is embedded in an $SU(3)$ subgroup of the full $R$-symmetry group, leading to $\cN=1$ supersymmetric models with $R$-symmetry $U(1)_R$. 
\item $\G$ is embedded in a specific $SU(2)$ subgroup of $SU(4)_R$, in which case the remaining supersymmetry is ${\cal N}=2$ with $R$-symmetry $SU(2)_R$. \end{enumerate}

The most plausible scenario for our purposes is the second, for a number of reasons. First and foremost, chiral theories can be defined only for ${\cal N}=0$ and ${\cal N}=1$ supersymmetry. Moreover, ${\cal N}=1$ supersymmetry guarantees the perturbative stability of the theory and also it addresses the infamous hierarchy problem. Henceforth we shall consider that $\G\subset SU(3)$. 

To proceed, it is useful to narrow the possibilities for the choice of discrete symmetry group $\G$. The most well-known cases 
include the discrete groups $\Z_{n}$ and $\Z_n\times \Z_m$. These cases were studied in the context of D-branes in \cite{Aldazabal:2000sa}. 
Here we shall focus on the case of $\Z_n$, and in particular in the most interesting case of $\Z_3$. 

The orbifold projection of the theory under consideration amounts to retaining only the invariant fields under the action of the 
discrete group. Therefore let us briefly describe the action of $\Z_3$ on the various fields of the theory. 
This action depends on their transformation properties under the $R$-symmetry and the gauge group. Let us consider a generator 
$g \in \Z_3$. This generator is conveniently labeled (see \cite{Douglas:1997de}) by three integers 
$\overrightarrow{a}\equiv (a_1,a_2,a_3)$ which satisfy the condition $a_1+a_2+a_3\equiv 0 \ \mbox{mod} \ 3$. 
We will focus on the case of $\cN=1$ with $a_i = (1, 1, -2)$.
For the gauge bosons the action is 
\be g\triangleright A_{\mu}=\g A_{\mu}\g^{-1}, 
\label{g-action-diag}
\ee  where \be\label{g3}
\g = \left(\begin{array}{ccc} \one_{N} & 0 & 0 \\
        0 & \omega\one_{N} & 0 \\
        0 & 0 & \omega^2\one_{N} \end{array}\right).
\ee Therefore the relevant projection is \be A_{\mu}=\g A_{\mu}\g^{-1}, \ee which means that the gauge group $SU(3N)$ of the original 
theory is broken down to $H=S(U(N)\times U(N)\times U(N))$ in the projected theory. 
The corresponding gauginos are singlets under \eq{g-action-diag} and have the same block-diagonal structure.
Let us note that in general the blocks of the 
matrix $\g$ could have different dimensionality 
(see e.g. \cite{Lawrence:1998ja,Aldazabal:2000sa,Kiritsis:2003mc}).
However, anomaly freedom of the projected theory typically
requires that the dimension of the three blocks is the same,
as will become obvious in the following. 

Now consider the complex scalars of the theory, i.e. the scalars $\f^i,i=1,2,3$, which correspond to the complexification of the six real ones of section 2. The action of the orbifold group $\Z_3$ is
\be 
g\triangleright \f^i=\o^{a_i}\g\f^i\g^{-1}, 
\label{orbifold-scalar}
\ee 
leading to the following projection condition,
\be \f^i_{IJ}=\o^{I-J+a_i}\f^i_{IJ}, \ee where $I,J$ are gauge indices. 
This means that $J=I+a_i$, so that the fields which survive the orbifold projection have the form $\f_{I,I+a_i}$ and 
 transform under the non-abelian factors of the gauge group $H$ as
\be
\label{repsH} 3\cdot \biggl((N,\overline{N},1)+(\overline{N},1,N)+(1,N,\overline{N})\biggl). 
\ee 
A similar condition holds for their fermionic partners, which must transform under 
$H$ in the representations (\ref{repsH}) exactly as the scalars. 
This is just another manifestation of the $\cN=1$ remnant supersymmetry. 
Moreover, the structure of the representations (\ref{repsH}) 
guarantees that the resulting theory does not suffer from any 
gauge anomalies\footnote{On the contrary, had we considered that 
the matrix (\ref{g3}) contained blocks of different dimensionality 
the projected theory would be anomalous and therefore additional 
sectors would be necessary in order to cancel the gauge anomalies.}. 

Let us next note two important features of the projected theory. First the fermions transform in chiral representations of the gauge group, the complex bifundamental representations (\ref{repsH}). Secondly, there are three fermionic generations in the theory, corresponding to the 3 chiral supermultiplets.

Concerning the interactions among the fields of the projected theory, let us consider the superpotential of the $\cN=4$ supersymmetric Yang-Mills theory, which has the form \cite{Brink:1976bc}:
\be\label{spot1} W_{\cN=4} =\epsilon_{ijk}Tr(\F^i\F^j\F^k)
,
\ee 
Here $\F^i$ denotes the three chiral superfields of the theory. 
Clearly, the superpotential after the orbifold projection has the same form but it encodes only the interactions among the surviving fields of the resulting ${\cal N}=1$ theory. Therefore it can be written as
\be\label{spot3} W_{{\cal N}=1}^{(proj)}=\sum_{I}\e_{ijk}\F^i_{I,I+a_i}\F^j_{I+a_i,I+a_i+a_j}\F^k_{I+a_i+a_j,I}
, \ee  
where the relation $a_1+a_2+a_3\equiv 0 \ \mbox{mod} \ 3$ was taken into account. 

>From the above superpotential one can easily read off the corresponding scalar (F-term) potential, which is
\be 
V_F(\f)= \frac 14 Tr([\f^i,\f^j]^\dagger [\f^i,\f^j]), 
\ee 
where $\f^i$ denotes the scalar component of the superfield $\F^i$. Moreover, there is a D-term contribution to the scalar potential, given by $V_{D}=\frac 12 D^2=\frac 12 D^{I}D_{I}$, where the $D$-terms have the form $D^{I}=\f_i^{\dagger}T^{I}\f^i$, where $T^I$ are the generators 
of the representation of the corresponding chiral multiplets.  The minimum of the full potential is obtained for vanishing vevs of the fields and therefore vacua corresponding to non-commutative geometries of the kind we are interested in do not exist without any additional modifications. Therefore at this stage we add ${\cal N}=1$ soft supersymmetry breaking 
(SSB) terms of the form\footnote{Here we present a set of scalar SSB terms. However, there exist of 
course other soft terms such as $\frac 12 M\l\l$, where $\l$ is the gaugino and $M$ its mass, 
which has to be included in the full SSB sector of a realistic theory \cite{Djouadi:2005gj}.}
\be\label{soft} 
V_{SSB}=\frac 12 \sum_i m^2_i\, {\f^i}^{\dagger}\f^i+\frac 12 \sum_{i,j,k}\, h_{ijk}\f^i\f^j\f^k+h.c., 
\ee
compatible with the orbifold group \eq{orbifold-scalar}.
Of course
a set of SSB terms in the potential is necessary anyway in order for the theory to have a 
chance to become realistic, see e.g. \cite{Djouadi:2005gj}. After the addition of these soft terms the 
full potential of the theory becomes 
\be\label{potential1} 
V=V_F+V_{SSB}+V_{D}.
\ee
This potential can be brought in the form
\be\label{potnewform} 
V=\frac 14 (F^{ij})^{\dag}F^{ij} \,\, + V_D 
\ee
for suitable parameters, where we have defined
\be
\label{fieldstrength} 
F^{ij}=[\f^i,\f^j]-i\e^{ijk}(\f^k)^{\dagger}.
\ee
Since the first term is positive definite, the global minimum of the potential is obtained when the following relations hold,
\bea
\label{twisted-vacuum} 
[\f^i,\f^j]&=&i\e_{ijk}(\f^k)^{\dagger}, \\ \f^i (\f^i)^{\dagger} &=&  R^2,
\eea
where $(\f^i)^{\dagger}$ denotes hermitean conjugation of the complex scalar field $\f^i$
and $[ R^2,\f^i] = 0$.
The above relations are closely related to a fuzzy sphere.
This can be seen by considering the untwisted fields $\tilde \f_i$, defined by
\be
\f^i = \Omega\,\tilde\f^i , 
\label{twisted-fields}
\ee
for some $\Omega\ne 1$ which satisfies
$
\Omega^3 = 1,  [\Omega,\f^i] = 0, \Omega^\dagger = \Omega^{-1}
\label{cond-1}
$
and\footnote{Here $[\Omega,\f^i]$ is understood before the orbifolding.}
$
(\tilde\f^i)^\dagger = \tilde\f^i, \mbox{i.e.}\,\,(\f^i)^\dagger = \Omega \f^i.
\label{cond-2}
$
Then \eq{twisted-vacuum} reduces to the ordinary fuzzy sphere relation
\be
\,[\tilde\f^i,\tilde\f^j] = i\e_{ijk}\tilde\f^k,
\label{fuzzy-transf} 
\ee
generated by $\tilde\f^i$, as well as to the relation
$
\tilde \f^i\tilde \f^i = R^2.
$
This justifies to call the noncommutative space generated by $\f^i$ a twisted fuzzy sphere,
denoted as $\tilde S^2_N$.
It is remarkable 
that this construction is possible only for $\Z_3$ and for no other
$\Z_n$, thus providing a justification for our choice of orbifold group.

It is quite straightforward to find configurations of $\f^i$ satisfying the relations of a twisted fuzzy sphere (\ref{twisted-vacuum}). 
Such a configuration is given by
\be
\label{solution1} 
\phi^i = \Omega\, (\one_3\otimes\lambda^i_{(N)}),
\ee 
where $\l^i_{(N)}$ denote the generators of $SU(2)$ in the $N$-dimensional irreducible representation and the matrix $\Omega$ is given by
\be
\Omega = \Omega_3 \otimes \one_N, \quad
\Omega_3 = \begin{pmatrix}
		0 & 1 & 0 \\
		0 & 0 & 1 \\
		1 & 0 & 0 \\
\end{pmatrix}, \quad \Omega_3^3 = \one. 
\label{omega-large}
\ee 
%
To understand the meaning of this configuration, it is helpful to consider a new basis 
where 
\be
\tilde\Omega_3 := U^{-1} \Omega_3 U = \diag(1,\omega,\omega^2)
\label{new-basis}
\ee 
is diagonal. 
Then  \eq{solution1} becomes
\be\label{transuntwist}
\f^i \, = \, \begin{pmatrix}
		\l^i_{(N)} & 0 & 0\\
		 0 & \omega\l^i_{(N)} & 0 \\
		0 & 0 & \omega^2\l^i_{(N)}  \\
\end{pmatrix},
\ee 
which can be interpreted as three identical fuzzy spheres (branes) embedded with relative angles $2\pi/3$.
This is clearly compatible with the orbifold constraint. 
The most general configurations in this basis would also contain specific off-diagonal matrices which connect these triple
spheres. This geometrical interpretation 
is helpful to understand the fluctuations around these solutions. 


The solution (\ref{solution1}) completely breaks the gauge symmetry $SU(N)^3$. 
However, for our purposes it will be useful to consider solutions which do not break the $SU(N)^3$ gauge symmetry completely but they break it down to a smaller gauge group. Such solutions are generically given by 
\be
\phi^i = \Omega\, \biggl(\one_3\otimes(\lambda^i_{(N-n)} \oplus 0_{n})\biggl),
\label{twistedfuzzys2-2}
\ee
where $0_{n}$ denotes the $n\times n$ matrix with vanishing entries. 
The gauge symmetry is broken from $SU(N)^3$ down to $SU(n)^3$. 
This vacuum should be interpreted as $\R^4 \times \tilde S^2_N$
with a twisted fuzzy sphere in the $\phi^i$ coordinates.

In order to understand the fluctuations of the scalar fields around 
this vacuum, the transformation \eq{new-basis} is useful.
Fluctuations around the ordinary fuzzy sphere 
$S^2_N$ are known to describe gauge and scalar 
fields on $S^2_N$ \cite{CarowWatamura:1998jn,Steinacker:2003sd}, 
which will turn into massive Kaluza-Klein modes
from the point of view of $\R^4$. More specifically,
we have seen in \eq{new-basis} 
that the twisted sphere $\tilde S^2_N$ is mapped via a unitary transformation $U$ into three identical fuzzy spheres
embedded in the diagonal 
$N\times N$ blocks of the original $3N\times 3N$ matrix. 
The vacuum can thus be interpreted at intermediate energy scales
as $\R^4 \times \tilde S^2_N$, or equivalently
three identical fuzzy spheres embedded with relative angles $2\pi/3$ in the $\phi^i$  coordinates.
Therefore all fluctuations can be understood as fields on these 
three diagonally embedded untwisted fuzzy spheres:
\be
\phi^i = \tilde \Omega (\lambda^i_{(N)} + A^i)
= \begin{pmatrix}
                \l^i_{(N)}+ A^i & 0 & 0\\
                 0 & \omega(\l^i_{(N)} + A^i) & 0 \\
                0 & 0 & \omega^2(\l^i_{(N)} + A^i) \\
\end{pmatrix},
\ee
as well as certain massive off-diagonal states which cyclically connect these spheres.
The field strength \eq{fieldstrength} reduces to the field strength on a 
fuzzy sphere 
\be
F^{ij}= [\f^i,\f^j]-i\e^{ijk}(\f^k)^{\dagger} 
\cong\, \tilde\Omega^2 ([\tilde\f^i,\tilde\f^j]-i\e^{ijk}\tilde\f^k) .
\ee
It now follows as in \cite{Aschieri:2006uw,Steinacker:2007ay}
that the block-diagonal bosonic and fermionic fields can be 
decomposed into Kaluza-Klein towers of massive modes
on $S^2_N$ resp. $\tilde S^2_N$
due to the Higgs effect. Similarly, the off-diagonal states 
connecting these branes are clearly massive, apart from would-be Goldstone bosons which are
absorbed in the massive gauge fields via the Higgs effect.
In generalized configurations such as \eq{twistedfuzzys2-2}, 
a massless sector remains which is chiral.

\section{An example: The trinification model}

In the previous section we presented a mechanism based on orbifolds and leading to chiral fermions in models with fuzzy extra dimensions. 
In the present section we apply the above ideas and we present an explicit example based on the gauge group $SU(3)^3$.

Let us therefore consider the case of $n=3$ in the notation of the previous section. Subsequently, let us consider the embedding
\be\label{dec3} SU(N)\supset SU(N-3)\times SU(3)\times U(1). \ee  Then the relevant embedding for the full gauge group is
\be S(U(N)^3)\supset SU(m)\times SU(3)\times SU(m)\times SU(3)\times SU(m)\times SU(3)\times U(1)^3\times U(1)^2. \ee 
where $m=N-3$.
The representations (\ref{repsH}) are then decomposed accordingly and as far as the non-abelian symmetry 
is concerned we obtain the following decomposition,
\bea \label{repsall}
&& SU(m)\times SU(m)\times SU(m)\times SU(3)\times SU(3)\times SU(3)\nn\\
&& (m,\obar{m},1;1,1,1)+(1,m,\obar{m};1,1,1)+(\obar{m},1,m;1,1,1)+\nn\\
&& +(1,1,1;3,\obar{3},1)+(1,1,1;1,3,\obar{3})+(1,1,1;\obar{3},1,3)+\nn\\
&& +(m,1,1;1,\obar{3},1)+(1,m,1;1,1,\obar{3})+(1,1,m;\obar{3},1,1)+\nn\\
&& +(\obar{m},1,1;1,1,3)+(1,\obar{m},1;3,1,1)+(1,1,\obar{m};1,3,1).
\eea
This is realized by the following vacuum, interpreted  
in terms of twisted fuzzy spheres $\tilde S^2_{m}$ 
as in \eq{twistedfuzzys2-2}:
\be
\phi^i = \Omega\,[\one_3\otimes (\lambda^i_{(m)} \oplus 0_{3})].
\label{twistedfuzzys2-4}
\ee 
Considering the decomposition (\ref{dec3}),
the gauge group is broken
to $K = SU(3)^3$. 
Finally, the surviving fields under the unbroken gauge group $K$ transform in the following representations,
\bea\label{repsfinal} & SU(3)\times SU(3)\times SU(3) \nn\\ 
		& 3\cdot \biggl((3,\overline{3},1)+(\overline{3},1,3)+(1,3,\overline{3})\biggl). \eea
These are the desired chiral representations of the trinification group $SU(3)_c\times SU(3)_L\times SU(3)_R$. This group was initially considered in \cite{Glashow:1984gc,Rizov:1981dp} and it was also studied in \cite{Leontaris:2005ax,Lazarides:1993uw,Babu:1985gi,Ma:2004mi} and from a string-theoretical perspective in \cite{Kim:2003cha} (see also \cite{ArkaniHamed:2001tb}). 
The quarks of the first family transform under the gauge group as
\begin{eqnarray}\label{quarks3}
q &=& \left(\begin{array}{ccc} d & u & h \cr d & u & h \cr d & u & h \end{array}\right)\sim (3,\overline 3,1), \\ 
q^c &=&\left(\begin{array}{ccc} d^c & d^c & d^c \cr u^c & u^c & u^c \cr h^c & h^c & h^c \end{array}\right) 
\sim (\overline 3,1,3),
\end{eqnarray}
and the leptons transform as
\begin{equation}\label{leptons3}
\lambda = \left(\begin{array}{ccc} N & E^c & \nu \cr E & N^c & e \cr \nu^c & e^c & S  \end{array}\right) 
\sim (1,3,\overline 3).
\end{equation}
Similarly, the corresponding matrices for the quarks and leptons of the other two families can be written down. Let us note that in the above matrices, along with the particles of the SM, new heavy quarks and leptons are accommodated.

The decomposition (\ref{repsall}) is very helpful in order to make two crucial remarks. First of all, it becomes clear 
from the vacuum solution (\ref{twistedfuzzys2-4}) that the scalar fields which acquire vevs in this vacuum are the following,
\be \langle(m,\obar{m},1;1,1,1)\rangle, \langle(1,m,\obar{m};1,1,1)\rangle, \langle(\obar{m},1,m;1,1,1)\rangle.
\ee
Then all the fermions, apart from the chirally protected ones, obtain masses, since we can form the invariants
\bea \label{inv1} && (1,m,\obar{m};1,1,1)\langle(m,\obar{m},1;1,1,1)\rangle (\obar{m},1,m;1,1,1)\qquad \mbox{+ cyclic permutations},\\  
\label{inv2} &&  (\obar{m},1,1;1,1,3)\langle(m,\obar{m},1;1,1,1)\rangle (1,m,1;1,1,\obar{3}) \qquad \mbox{etc.,}
\eea 
and the corresponding ones for all the other fermions. In these invariants the field in the middle is the scalar field which acquires the vev \eq{twistedfuzzys2-4}, 
while the other two are fermions, i.e. the invariants are trilinear Yukawa terms and they are responsible for the 
fermion masses after the spontaneous symmetry breaking. Therefore a finite Kaluza-Klein tower of massive fermionic modes appears, 
consistent with the interpretation of the vacuum (\ref{twistedfuzzys2-4}) as a higher-dimensional theory 
with spontaneously generated fuzzy extra dimensions. In particular, the fluctuations from this vacuum correspond to 
the internal components of the higher dimensional gauge field. 
On the other hand, for the fermions transforming as 
$(1,1,1;3,\overline 3,1),(1,1,1;\overline 3,1,3)$ and $(1,1,1;1,3,\overline 3)$ there exists no
trilinear invariant that could be formed with one of the scalar fields which acquire a vev. 
Therefore they remain massless, and these are the low-energy chiral fermions of the model. 

Once the supersymmetric trinification model is obtained, the next step is to study its further breaking down to the MSSM 
and to the $SU(3)\times U(1)_{em}$. One way to perform this task is to treat the model as an ordinary
Grand Unified Theory (GUT)  and proceed to its spontaneous symmetry breaking \cite{Babu:1985gi,Lazarides:1993uw,Ma:2004mi}.
 Here we would like to discuss an alternative procedure, which is based on the mechanism of section 3.
Let us consider the superpotential of the $SU(3)^3$ model, which has the form
\be
W= \e_{ijk} Tr(Q^i(Q^c)^j\L^k),
\ee
where the index $i=1,2,3$ counts the three families and $Q,Q^c,\L$ are the superfields
corresponding to $q,q^c,\l$ respectively. Two remarks are in order. First, the above superpotential is not 
sufficient in order to apply our mechanism.
This is evident from the fact that the scalar components of $Q$ and $Q^c$ cannot acquire a vev, since such a vev would break the
colour group and QED. Therefore, only the scalar component of $\L$ may obtain a vev, and no matter which soft terms are added  
the above superpotential cannot lead to dynamical generation of fuzzy extra dimensions.
Secondly, the above superpotential is not the most general one invariant under all the symmetries of
 the $SU(3)^3$ model. It is just the superpotential which arises after the orbifold projection of the
 initial ${\cal N}=4$ SYM theory and the subsequent dynamical generation of twisted fuzzy spheres.
However, the most general invariant superpotential of the model has the following form
\be
\label{spot4}
 W=Y_{ijk}Tr(Q^i(Q^c)^j\L^k)
+Y'_{ijk}Tr(\L^i\L^j\L^k+(Q^c)^i(Q^c)^j(Q^c)^k
+Q^iQ^jQ^k),
\ee
 where $Y_{ijk}$ and $Y'_{ijk}$ have in general a symmetric and an antisymmetric part.
These two parts generate interactions within the same family and interactions mixing the families respectively.
>From the point of view of the ${\cal N}=4$ theory, the extra terms appearing in (\ref{spot4}) may arise from marginal and relevant deformations of ${\cal N}=4$ SYM \cite{Leigh:1995ep}.

Let us therefore consider the superpotential (\ref{spot4}), which involves a term proportional to
$Tr(\L\L\L)$. This term is important since the scalar components of each $\L^i$
may acquire a vev. Of course, these vevs cannot appear in any entry of the corresponding $3\times 3$ matrices but only in the entries with vanishing charge. Referring to (\ref{leptons3}), the entries of the corresponding scalar matrix which may acquire a vev are the{\footnote{Tilded fields denote the
 scalar superpartners of the corresponding fermions.}
$\tilde\nu,\tilde\nu^c,\tilde S$ (which are responsible for the GUT breaking) and the
$\tilde N,\tilde N^c$ (which are responsible for the electroweak breaking). In \cite{Chatzistavrakidis:2010xi} it was shown that
in principle there exist vacua of this model where the breaking of $SU(3)^3$ down to the MSSM and the $SU(3)\times U(1)_{em}$
 occurs and twisted fuzzy spheres are dynamically generated. In this picture the breaking of the trinification model
acquires an interesting geometrical explanation in terms of dynamically generated fuzzy extra dimensions. 

\section{Conclusions}

In the present article we exhibit various models of 4D Yang-Mills gauge theory, which at intermediate energies 
behaves as higher-dimensional theory with effective geometry $M^4 \times K$. Here $K$ can be a compact fuzzy space
such as $S^2_N$, $S^2_N \times S^2_N$, or similar spaces. This fuzzy extra-dimensional space 
arises through nontrivial expectation values of the nonabelian scalar fields in the adjoint of the gauge group,
via the usual Higgs mechanism. One recovers indeed the expected tower of Kaluza-Klein states, which is  truncated
at high energies.  The model is then effectively described as a higher-dimensional gauge theory,
and equivalently as a 4D renormalizable gauge theory with large gauge group.

However, in the simplest examples the low-energy effective theory turns out to be 
non-chiral. 
In order to obtain a chiral low-energy theory relevant to particle physics, we therefore consider orbifold models.
The starting point is a $\cN=1$ supersymmetric gauge theory obtained by restricting $\cN=4$ SYM to the invariant 
sector under a $\Z_3$ group action.  Adding a particular set of soft supersymmetry breaking terms we show that this model 
can also develop fuzzy extra dimensions, leading to a chiral low-energy theory.
More precisely, we identify vacua which can be interpreted as twisted fuzzy sphere, or equivalently three identical fuzzy spheres 
embedded with a relative angle of $2\pi/3$.
Again, a tower of Kaluza-Klein states is found. Furthermore, we give an example of a model which gives
an extension of the standard model in its low-energy sector.

The examples presented here are very basic, and there is plenty of room for variations and generalizations.
In particular, the orbifold setting is very attractive from the physics point of view, and there is hope
than one may obtain candidates for realistic models in this way. These constructions are also very natural from the 
string theory point of view, since they lead to higher-dimensional geometrical structures which are 
very familiar from string theory. However, it is very remarkable that these higher-dimensional structures arise 
here within renormalizable and possibly even finite 4D gauge theory models. Finally,
the relation with string theory also suggests to generalize these models in order to reproduce other geometrical
structures encountered in string theory. A parallel approach would be the study of matrix models for string/M-theory \cite{Ishibashi:1996xs,Banks:1996vh} in the above spirit{\footnote{For related work see e.g. \cite{Aoki:2002jt,Itoyama:1997gm}}. These are directions for future work.

\paragraph{Acknowledgments}

The pleasant and stimulating atmosphere at the 2010 Corfu Summer 
Institute of elementary particle physics provided an opportunity for exchange of ideas and discussions, which is gratefully 
acknowledged. This work was partially supported
by the European Union 7th network program ``Unification in the LHC era'' (PITN-GA-2009-237920) and the NTUA's programs supporting basic
research PEBE 2009 and 2010.
The work of H.S. was supported by the FWF project  P21610.
G.Z. would like to thank the ITP
Heidelberg for the very warm hospitality.


\begin{thebibliography}{99}

\bibitem{Antoniadis:1990ew}
  I.~Antoniadis,
  {\it A Possible new dimension at a few TeV,
  Phys.\ Lett.\ }  {\bf B246 } (1990)  377-384.

\bibitem{Seiberg:1999vs}
  N.~Seiberg and E.~Witten,
  {\it String theory and noncommutative geometry,
  JHEP } {\bf 9909} (1999) 032
  [arXiv:hep-th/9908142].

\bibitem{Hull:1994ys}
  C.~M.~Hull, P.~K.~Townsend,
  {\it Unity of superstring dualities,
  Nucl.\ Phys.\ } {\bf B438 } (1995)  109-137.
  [hep-th/9410167].

\bibitem{Ishibashi:1996xs}
  N.~Ishibashi, H.~Kawai, Y.~Kitazawa {\it et al.},
  {\it A Large N reduced model as superstring,
  Nucl.\ Phys.\  }{\bf B498}, 467-491 (1997).
  [hep-th/9612115].

\bibitem{Chatzistavrakidis:2010tq}
  A.~Chatzistavrakidis, G.~Zoupanos,
  {\it Higher-Dimensional Unified Theories with Fuzzy Extra Dimensions,
  SIGMA }{\bf 6 } (2010)  063.
  [arXiv:1008.2049 [hep-th]].

\bibitem{Aschieri:2003vy}
  P.~Aschieri, J.~Madore, P.~Manousselis and G.~Zoupanos,
  {\it Dimensional reduction over fuzzy coset spaces,
  JHEP } {\bf 0404}, 034 (2004)
  [arXiv:hep-th/0310072].


\bibitem{Aschieri:2006uw}
  P.~Aschieri, T.~Grammatikopoulos, H.~Steinacker {\it et al.},
  {\it Dynamical generation of fuzzy extra dimensions, dimensional reduction and symmetry breaking,
  JHEP } {\bf 0609 } (2006)  026. [hep-th/0606021].

\bibitem{Chatzistavrakidis:2010xi}
  A.~Chatzistavrakidis, H.~Steinacker, G.~Zoupanos,
  {\it Orbifolds, fuzzy spheres and chiral fermions,
  JHEP } {\bf 1005 } (2010)  100.
  [arXiv:1002.2606 [hep-th]].

\bibitem{Brink:1976bc}
  L.~Brink, J.~H.~Schwarz and J.~Scherk,
  {\it Supersymmetric Yang-Mills Theories,
  Nucl.\ Phys.\  } B{\bf 121} (1977) 77;
  F.~Gliozzi, J.~Scherk and D.~I.~Olive,
  {\it Supersymmetry, Supergravity Theories And The Dual Spinor Model,
  Nucl.\ Phys.\ } B{\bf 122} (1977) 253.

\bibitem{Steinacker:2003sd}
  H.~Steinacker,
  {\it Quantized gauge theory on the fuzzy sphere as random matrix model,
  Nucl.\ Phys.\  } {\bf B679 } (2004)  66-98.
  [hep-th/0307075].


\bibitem{Steinacker:2007ay}
  H.~Steinacker and G.~Zoupanos,
 {\it Fermions on spontaneously generated spherical extra dimensions,
  JHEP } {\bf 0709} (2007) 017
  [arXiv:0706.0398 [hep-th]].

\bibitem{Chatzistavrakidis:2009ix}
  A.~Chatzistavrakidis, H.~Steinacker, G.~Zoupanos,
  {\it On the fermion spectrum of spontaneously generated fuzzy extra dimensions with fluxes,
  Fortsch.\ Phys.\  } {\bf 58 } (2010)  537-552.
  [arXiv:0909.5559 [hep-th]].


\bibitem{Uranga:2002vk}
  A.~M.~Uranga,
  {\it D-branes, fluxes and chirality,
  JHEP } {\bf 0204 } (2002)  016.
  [hep-th/0201221].

\bibitem{Douglas:1997de}
  M.~R.~Douglas, B.~R.~Greene, D.~R.~Morrison,
  {\it Orbifold resolution by D-branes,
  Nucl.\ Phys.\  } {\bf B506 } (1997)  84-106.
  [hep-th/9704151].

\bibitem{Myers:1999ps}
  R.~C.~Myers,
  {\it Dielectric branes,
  JHEP } {\bf 9912 } (1999)  022.
  [hep-th/9910053].

\bibitem{Dixon:1985jw}
  L.~J.~Dixon, J.~A.~Harvey, C.~Vafa {\it et al.},
  {\it Strings on Orbifolds,
  Nucl.\ Phys.\  }{\bf B261 } (1985)  678-686;
  {\it Strings on Orbifolds. 2.,
  Nucl.\ Phys.\  }{\bf B274 } (1986)  285-314.
  

\bibitem{Kachru:1998ys}
  S.~Kachru, E.~Silverstein,
  {\it 4-D conformal theories and strings on orbifolds,
  Phys.\ Rev.\ Lett.\  } {\bf 80 } (1998)  4855-4858.
  [hep-th/9802183].

\bibitem{Douglas:1996sw}
  M.~R.~Douglas, G.~W.~Moore,
  {\it D-branes, quivers, and ALE instantons,}
  [hep-th/9603167].

\bibitem{Aldazabal:2000sa}
  G.~Aldazabal, L.~E.~Ibanez, F.~Quevedo {\it et al.},
  {\it D-branes at singularities: A Bottom up approach to the string embedding of the standard model,
  JHEP }{\bf 0008 } (2000)  002.
  [hep-th/0005067].

\bibitem{Lawrence:1998ja}
  A.~E.~Lawrence, N.~Nekrasov and C.~Vafa,
  {\it On conformal field theories in four dimensions,
  Nucl.\ Phys.\  } B{\bf 533}, 199 (1998)
  [arXiv:hep-th/9803015].

\bibitem{Kiritsis:2003mc}
  E.~Kiritsis,
  {\it D-branes in standard model building, gravity and cosmology,
  Fortsch.\ Phys.\  } {\bf 52} (2004) 200
  [{\it Phys.\ Rept.\  }{\bf 421} (2005\ ERRAT,429,121-122.2006) 105]
  [arXiv:hep-th/0310001].

\bibitem{Djouadi:2005gj}
  A.~Djouadi,
  {\it The Anatomy of electro-weak symmetry breaking. II. The Higgs bosons in the
  minimal supersymmetric model,
  Phys.\ Rept.\  } {\bf 459}, 1 (2008)
  [arXiv:hep-ph/0503173].


\bibitem{CarowWatamura:1998jn}
  U.~Carow-Watamura and S.~Watamura,
  {\it Noncommutative geometry and gauge theory on fuzzy sphere,
  Commun.\ Math.\ Phys.\ } {\bf 212} (2000) 395
  [arXiv:hep-th/9801195].



\bibitem{Glashow:1984gc}
  S.~L.~Glashow,
  {\it Trinification Of All Elementary Particle Forces,}
in Fifth Workshop on Grand Unification edited by K.
Kang, H. Fried and F. Frampton (World Scientific, Singapore, 1984). p.
88.

\bibitem{Rizov:1981dp}
  V.~A.~Rizov,
  {\it A Gauge Model Of The Electroweak And Strong Interactions Based On The Group
  $SU(3)_L \times SU(3)_R \times SU(3)_C$,
  Bulg.\ J.\ Phys.\ } {\bf 8}, 461 (1981).


\bibitem{Babu:1985gi}
  K.~S.~Babu, X.~G.~He and S.~Pakvasa,
  {\it Neutrino Masses And Proton Decay Modes In $SU(3) \times SU(3) \times SU(3)$
 Trinification,
  Phys.\ Rev.\  } D {\bf 33} (1986) 763.

\bibitem{Lazarides:1993uw}
  G.~Lazarides and C.~Panagiotakopoulos,
  {\it MSSM from SUSY trinification,
  Phys.\ Lett.\  } B{\bf 336}, 190 (1994)
  [arXiv:hep-ph/9403317].


\bibitem{Ma:2004mi}
  E.~Ma, M.~Mondragon and G.~Zoupanos,
  {\it Finite $SU(N)^k$ unification,
  JHEP } {\bf 0412}, 026 (2004)
  [arXiv:hep-ph/0407236];




\bibitem{Leontaris:2005ax}
  G.~K.~Leontaris and J.~Rizos,
  {\it A D-brane inspired $U(3)_C \times U(3)_L \times U(3)_R$ model,
  Phys.\ Lett.\  } B{\bf 632} (2006) 710
  [arXiv:hep-ph/0510230].

\bibitem{Kim:2003cha}
  J.~E.~Kim,
  {\it Z(3) orbifold construction of $SU(3)^3$ GUT with $sin^2(\theta(0)(W)) =
  3/8$,
  Phys.\ Lett.\  } B{\bf 564}, 35 (2003)
  [arXiv:hep-th/0301177].
  K.~S.~Choi and J.~E.~Kim,
  {\it Three family Z(3) orbifold trinification, MSSM and doublet-triplet
  splitting problem,
  Phys.\ Lett.\  } B{\bf 567}, 87 (2003)
  [arXiv:hep-ph/0305002].

\bibitem{ArkaniHamed:2001tb}
  N.~Arkani-Hamed, T.~Gregoire, J.~G.~Wacker,
  {\it Higher dimensional supersymmetry in 4-D superspace,
  JHEP } {\bf 0203 } (2002)  055.
  [hep-th/0101233].

\bibitem{Leigh:1995ep}
  R.~G.~Leigh, M.~J.~Strassler,
  {\it Exactly marginal operators and duality in four-dimensional N=1 supersymmetric gauge theory,
  Nucl.\ Phys.\  }{\bf B447 } (1995)  95-136.
  [hep-th/9503121].


\bibitem{Banks:1996vh}
  T.~Banks, W.~Fischler, S.~H.~Shenker {\it et al.},
  {\it M theory as a matrix model: A Conjecture,
  Phys.\ Rev.\  }{\bf D55 } (1997)  5112-5128.
  [hep-th/9610043].

\bibitem{Aoki:2002jt}
  H.~Aoki, S.~Iso, T.~Suyama,
  {\it Orbifold matrix model,
  Nucl.\ Phys.\  }{\bf B634 } (2002)  71-89.
  [hep-th/0203277].

\bibitem{Itoyama:1997gm}
  H.~Itoyama, A.~Tokura,
  {\it USp(2k) matrix model: F Theory connection,
  Prog.\ Theor.\ Phys.\  }{\bf 99 } (1998)  129-138.
  [hep-th/9708123];
  {\it USp(2k) matrix model: Nonperturbative approach to orientifolds,
  Phys.\ Rev.\  }{\bf D58 } (1998)  026002.
  [hep-th/9801084].

\end{thebibliography}
\end{document}